\documentclass[12pt]{iopart}

\usepackage{graphicx} 
\usepackage{bbm}
\usepackage{amssymb}                           
\usepackage{cite}
          
\begin{document}

\title[Random topologies and the emergence of cooperation: the role of short-cuts]{Random topologies and the emergence of cooperation: the role of short-cuts}

\author{D. Vilone$^{1,*}$, A. S\'anchez$^{1,2,3,**}$, and J. G\'omez-Garde\~nes$^{3,4,***}$}
\address{$^1$ Grupo Interdisciplinar de Sistemas Complejos (GISC), Departamento de Matem\'aticas, Universidad
Carlos III de Madrid, 28911 Legan\'es, Madrid, Spain}
\address{$^2$ Instituto de Ciencias Matem\'aticas CSIC-UAM-UC3M-UCM, 28049 
Cantoblanco, Madrid, Spain}
\address{$^3$ Instituto de Biocomputaci\'on y F\'\i sica de Sistemas Complejos,
Universidad de Zaragoza,
Campus R\'\i o Ebro, 50018 Zaragoza, Spain}
\address{$^4$ Departamento de Matem\'atica Aplicada, Universidad Rey Juan Carlos, 
28933 M\'ostoles, Madrid, Spain}
\address{$^*$ E-mail: daniele.vilone@gmail.com}
\address{$^{**}$ E-mail: anxo@math.uc3m.es}
\address{$^{***}$ E-mail: gardenes@gmail.com}

\begin{abstract}
We study in detail the role of short-cuts in promoting the emergence of cooperation 
in a network of
agents playing the Prisoner's Dilemma Game (PDG). We introduce a model
whose topology interpolates between the
one-dimensional euclidean lattice (a ring) and the complete graph by
changing the value of one parameter (the probability $p$ to add a link
between two nodes not already connected in the euclidean
configuration). We show that there is a region of values of $p$ in
which cooperation is largely enhanced, whilst for smaller values of
$p$ only a few cooperators are present in the final state, and for
$p\rightarrow1^-$ cooperation is totally suppressed. We present
analytical arguments that provide a very plausible interpretation of
the simulation results, thus unveiling the mechanism by which 
short-cuts contribute to promote (or suppress) cooperation.
\end{abstract}


\section{Introduction}
\label{intro}

Cooperative behaviours are commonly observed in nature and in human
society. However, explaining its origin is not a trivial
task~\cite{Darwin,Axel81,Mayn95}. Indeed, at the level of individuals,
selfish attitudes are often more convenient (as can be easily
seen in many animal behaviours), so that there must be some
mechanisms that promote a sort of crossover from the ``micro"
level ({\em i.e.}, in the interactions between a few individuals), to
the ``macro" one (when a great number of individuals is involved),
where cooperation in many cases prevails over defection, reversing the
microscopic selfish trend.

Already Nowak and May~\cite{Now92} suggested that the topology of
interactions could be a fundamental factor enhancing the emergence of
cooperation even though the interaction at the individual level. 
A wealth of studies on evolutionary game theory on graphs
spawned from this first, key insight \cite{sza,tomassini,roc09},
making clear that many factors can favour (or hinder) global
cooperative behaviours, as for instance the update rule the
individuals use to evolve their strategies (see Section~\ref{rslts}
below) or the details of the topology; in other words, the outcome 
of evolutionary games on graphs is far from being universal. From 
the experimental viewpoint, the situation is similar, and reports that
the topology influences \cite{cassar} or does not influence \cite{watts}
the emergence of cooperation have been published (although 
it must be kept in mind the 
networks used in those works are very small, see \cite{helbing}, the 
maximum size ever tried being 13$\times$13 \cite{us}).  Therefore, it is
worth improving our understanding of the role of both topology and
update rules in enhancing (or preventing) cooperation.

Among the topological features that
play a key role in evolutionary games, prominent ones are the
degree distribution and the
clustering. 
The degree distribution
and in particular whether it is of exponential type or has a long tail
(scale free in general) has been shown to lead to dramatical
differences in behaviour \cite{sp,gomezgardenes}. As for the
clustering, research indicates that,
depending on the specific dynamics, 
highly clustered networks can promote the survival of
initially cooperating agents as they interact mainly among themselves,
subsequently fostering the spreading of their strategy throughout the
whole system~\cite{Now92,roc09,asse,pusch}.  This article is intended
to disentagle these effects of clustering from those arising from the 
existence of short-cuts in the network, which can coexist in many
relevant situations. Thus, the
Watts-Strogats (WS) small-world \cite{ws98} network is 
the paradigmatic example of local behaviour, 
like an
euclidean lattice, while acting globally like a random network (thus having high
clustering coefficient and low diameter). Therefore, in
this paper we will focus on a model network inspired by the WS one,
but with some remarkable differences with respect to it,
in order to gain insights on the role of
short-cuts in the emergence and sustainability of cooperation.

To this end, the article has the following structure: In Section~\ref{mdl} 
we summarize the background on the issue we address here. 
Subsequently, in Section~\ref{swn} we introduce
the model network which we use as the tool of the study,
basically a small-world network with additional links. 
Section~\ref{rslts} presents then our main simulation results,
followed in Section~\ref{intrprt} by some theoretical interpretations
and implications. Finally, in Section~\ref{cncl} we summarize our
conclusions and discuss the future perspectives they open.

\section{The Prisoner's Dilemma Game and the evolution of cooperation}
\label{mdl}

In a general evolutionary framework, $N$ individuals interact through
a game. At the initial step every individual plays with a subset of
the players and receives a total payoff according to her action and
her partners' one. After that, players update their strategy according
to an {\it a priori} specified update rule. Within this framework, to
properly specify an evolutionary game one has to choose a game
(defined by the payoffs to the different actions players can take), an
interaction set (with whom every agent plays each time) and a
payoff-dependent update rule (how agents modify the strategy governing
their actions). Evolutionary game theory on graphs~\cite{sza} arises
when the interaction set is defined as a network: the subset of the
players with whom a given one interacts is the set of her
neighbours. Thus, in a complete graph all individuals are connected
(well-mixed population case or limit); in an euclidean square
bidimensional lattice every agent interacts with its four nearest
neighbours, and so on. On a complex network, the structure can be very
complicated and the number of neighbours is in general different for
each agent.

The game we will consider is the Prisoner's Dilemma game (PDG), the
paradigmatic model to study the emergence of cooperation
\cite{Axel81}.  The PDG is a $2\times2$ symmetric game; each player
can cooperate (C), or defect (D). If both players cooperate, they get
a reward $R$, if both defect they get instead a punishment $P$, and
finally if they adopt different strategies the defector earns the
temptation payoff $T$, whilst the cooperator receives the sucker's one
$S$, with $T>R>P\geq S$.  Therefore, the unique Nash equilibrium (a
choice of actions for the players from which none of them has
incentives to deviate, see, e.g., \cite{Hof98}) is mutual defection,
even though this is not a Pareto optimal solution, {\em i.e.}, there is
another configuration, mutual cooperation, in which both players would
earn more: here lies the dilemma.  Following Nowak and
May~\cite{Now92}, who first show that cooperation could be enhanced by
the presence of a lattice in the PDG, we choose $R=1$, $T=b\in (1,2]$
  and $P=S=0$. This choice (often referred to as the ``weak PDG") does
  not change the essential physics of the model, makes the study
  easier and reduces to one the number of free parameters of the game.
  While in a well-mixed population (complete graph) a system of agents
  playing the PDG ends up in a frozen state with only defectors
  \cite{Hof98,per09,sza,moya} independently of the particular update
  rule adopted by the agents, when the population structure is a
  lattice, the system may converge to a final state with a non
  vanishing density of survived
  cooperators~\cite{Now92,roc09,sza,moya,per9}. As mentioned above,
  subsequent research showed that the effect of the topology is not
  universal, mainly because it also depends strongly on the update
  rule adopted by the players~\cite{roc09,per9,hau,sys,oht} and not
  only on the particular topology. Below (see Sec.\ \ref{rslts}) we
  will consider two update rules in order to asses the scope of our
  findings.
    
Regarding the second ingredient of the game, the network, our departure point is 
the WS one, in its two
different versions, WS \cite{ws98}, and Newman--Watts~\cite{nw}
(NW). In the first one each link connecting two nearest neghbours in an unidimensional
ring is rewired (that is, one of its two nodes is changed) with probability $p$, whereas in the latter
case a short-cut (a new link connecting two originally separated nodes) is added to
each site again with probability $p$. Both versions of the model have the
property that, according to a parameter $p\in[0,1]$, its topology can be tuned from
the euclidean (for $p=0$) to the completely random one ($p=1$). In the middle, there
is a range of values (approximatedly $0<p\lesssim0.1$) for which the network locally
still behaves like a regular lattice, but globally already as a random network~\cite{ws98,nw,bw00}.
Such models allow us to focus on the effect the short-cuts in a complex network have
over the dynamics and evolution of cooperative behaviours.

To our knowledge, the fact that inserting long-range connections into
a regular ring can sensitively influence the evolution and stability
of global cooperative behaviours was first noticed in
\cite{masu}. Subsequently, researchers started studying the specific
influence of the decrease of the diameter due to the short-cuts and
the onset of heterogeneity in the network (that is, the appearance of
hubs created by the added links). Thus, Santos et al.~\cite{san05}
showed that in a perfectly homogeneous WS network cooperation is
actually enhanced, but less than in a heterogeneous network.  Fu et
al.~\cite{fu07} reported that there is a peak of the final cooperator
density at a certain value of degree heterogeneity, working with a NW
network where the short-cuts are added to some fixed hubs in order to
have a specified degree of heterogeneity $h$, defined as
\begin{equation}
\label{het}
h=\frac{1}{N}\sum_k k^2n(k)-\langle k\rangle^2
\end{equation}
where $N$ is the number of nodes of the network, $n(k)$ gives the
number of vertices with $k$ edges and $\langle k\rangle$ is the
average degree. The maximum value of the final cooperation is reached
around $h\simeq0.2$. A similar result was reached by Du {\em et
  al.}\ \cite{du09}.  Additionally, Vukov et al~\cite{vuk08}, showed
that for homogeneous networks with dynamical noise (that is, with the
possibility for an individual to adopt the strategy of worse
performing neghbours) long-range connections can diminish the final
defector density.  Other studies on small world (or similar) networks,
but utilizing different games~\cite{wu05,ren07,du10} or different
details of the dynamics, confirmed that short-cuts have a non trivial
and often favourable effect on the emergence of cooperation.  On the
other hand, on an Erd\"os-R\'enyi (ER) network (totally random
topology), cooperation is much more enhanced that in both regular
lattices and small-world systems~\cite{cardillo}, a result compatible
with the previous ones. In view of all this research, we can now state
that tuning a network from the one-dimensional ring to a totally
random network, the cooperation in the final state increases in
general, even though there can be factors in the dynamics which could
sometimes alter this picture. The question then arises naturally as to what
occurs if we go instead from the ring topology to the complete graph,
in which we know that cooperation is totally suppressed.  This is the
specific question we aim at clarifying in this work.

\section{The Link Added Small World (LASW) network}
\label{swn}

Let us now introduce the model for the Link Added Small World (LASW, in the following)
network. The LASW network is defined from a regular one-dimensional
ring with $N$ nodes: each site is connected to its $2m$ nearest
neighbours ($m\in\mathbb{N}$), so that there are $mN$ links (or
edges). The diameter $D_0$ ({\em i.e.}, the average distance between
two randomly chosen nodes) and the clustering coefficient $\chi_0(m)$
({\em i.e.}, the probability that two neighbours of a third one are
also neighbours of each other) of such a simple graph are~\cite{bw00}
\begin{equation}
\label{D0} D_0=\frac{N}{2m}\ ,
\end{equation}
\begin{equation}
\label{chi0} \chi_0(m)=\frac{3(m-1)}{2(2m-1)}\ .
\end{equation}
At this point, we can modify the regular ring by adding new links;
that is, we add each of the $[N(N-1)/2-mN]$ missing edges with
probability $p$. Therefore, by varying $p$ we can tune the topology
from euclidean to the complete-network as desired.  In this model
network, it is easy to determine that for this kind of graphs the
diameter is given by
\begin{equation}
\label{D_la}
D=D_0\cdot f[(N-2m-1)p]\ ,
\end{equation}
where the quantity $(N-2m-1)p=x$ is the density of short-cuts per site
and
\begin{equation}
\label{F_la} f(x)= \left\{
\begin{array}{ll}
1 & \ \ \ \ \ \ \ \ \ \ \ \ \ \ \ \ \ \ x\rightarrow 0^+ \\
\ \ & \ \ \\
\frac{\log (x)}{x} & \ \ \ \ \ \ \ \ \ \ \ \ \ \ \ \ \ \
x\gg 1\ .
\end{array}
\right.
\end{equation}
At the same time, the clustering coefficient is given in the
thermodynamical limit by
\begin{equation}
\label{chi_la} \chi(p,m)= \left\{
\begin{array}{ll}
\chi_0(m) & \ \ \ \ \ \ \ \ \ \ \ \ \ \ \ \ \ \ p=0 \\
\ \ & \ \ \\
p & \ \ \ \ \ \ \ \ \ \ \ \ \ \ \ \ \ \ p\in(0,1]\ .
\end{array}
\right.
\end{equation}
Notice how for $p\neq0$ the clustering coefficient is exactly given by
the link adding probability. The previous expression can be found
directly by taking the thermodynamic limit $N\rightarrow+\infty$ in
the general result for the clustering coefficient
\[
\chi_N(p,m)= z\cdot\left[\chi_0(m)+[1-\chi_0(m)]\cdot p\right]+(1-z)\cdot p\ ,
\]
where $z=z(N,m,p)=2m/\bar{n}(N,m,p)$ is the probability that two nodes
are not neighbours in the one-dimensional ring, and
$\bar{n}=2m+p(N-1-2m)$ is the average number of neighbours per
site. We highlight that, except for the case $m=1$, the sequence
of functions $\chi_N(p,m)$, {\it i.e.} the sequence of cluster
coefficients given by different values of $N$, does not converge
uniformly to the expression of $\chi(p,m)$ given in
Eq.~(\ref{chi_la}), which is, therefore, a non-continuous function in
$p$. On the other hand, such pathological behaviour is not alarming,
since simulations never deal with really infinite systems.

It is worth highlighting some remarkable topological features, which
will be useful below.  First of all, let us consider for simplicity
the case $N\gg2m+1$, that is $N-2m-1\simeq L$.  Calling $p^*=2m/N$, we
have a crossover in the behaviour of a LASW network. For $p\ll p^*$
the number of short-cuts is still much smaller than the number of
regular links, and the system behaves similarly to a WS network with
rewiring probability $p_r=Np/(2m)$.  In the opposite limit $p\gg p^*$,
the number of added links is much larger than that of regular ones,
and thus the behaviour of the system approach that of an ER
network~\cite{erdr}.  Thus, a LASW network is in practice the sum of a
regular graph and an ER one, so that its degree distribution $P(k)$ is
always equal to $P_{ER}(k+2m)$, $P_{ER}(k)$ being the equivalent
distribution of an ER network with the same $p$ and $N$ [{\em i.e.} a
  binomial distribution with mean value $pN$ and variance $p(1-p)N$].
In the thermodynamical limit ($N\rightarrow+\infty$) we have
$p^*\rightarrow0^+$ and the LASW network behaves exactly like an ER
one $\forall p\in(0,1]$, with neither its diameter nor its clustering
  coefficient depending on $m$ [except for
    $\chi(p=0,m)=\chi_0(m)$]. Moreover, the clustering coefficient for
  $m\geq 2$ becomes equal to the value it has exactly for $m=1$. In
  practice, for very large networks, the topology is independent of
  $m$.

\section{Prisoner's Dilemma game on LASW networks: Simulation results}
\label{rslts}

\begin{figure}
\begin{center}
\includegraphics[angle=0,width=9.1cm,clip]{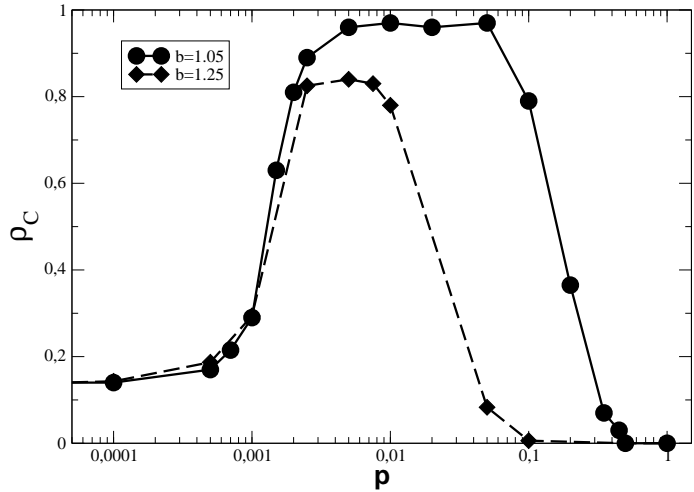}
\end{center}
\caption{Cooperator density $\rho_C$ [for temptation values of
    $b=1.05$ (circles) and $b=1.25$ (diamonds)] in the asymptotic
  state for LASW networks with $m=1$ and $N=1000$ as a function of the
  link-adding probability $p$. All agents use the UI update rule. Note
  the logarithmic scale of the $x$ axis.}
\label{lak1}
\end{figure}

We now use our model to investigate how the topology influences the
emergence of cooperation in a population whose individuals play the
PDG.  As stated above, we still need to introduce the strategy update
rules. Even though there are many possible choices in
literature~\cite{roc09,per09,sza}, we present here only the two most
commonly considered ones: Unconditional imitation (UI) and replicator
rule (REP). UI is a completely deterministic rule (an automaton): At
the end of each round of the game, every player imitates the strategy
of the neighbour which has obtained the best payoff provided it is
larger that her own payoff. In REP, agents choose a neighbour at
random: If the payoff of the chosen neighbour is lower than the
agent's own, nothing happens. If it is larger, the agent will adopt
the neighbour's strategy with a probability proportional to the
difference between the two payoffs. These two rules differ in
character (deterministic {\it versus} stochastic) and in their
outcomes in the presence of a homogeneous network structure: indeed,
UI promotes very much the appearance of cooperation whereas REP does
not~\cite{Now92,roc09,per9,hau}. Therefore, while these two rules are
by no means the whole panorama of possible dynamics, they will allow
us to compare different situations as far as the emergence of
cooperation is concerned.

\begin{figure}
\begin{center}
\includegraphics[angle=0,width=9.1cm,clip]{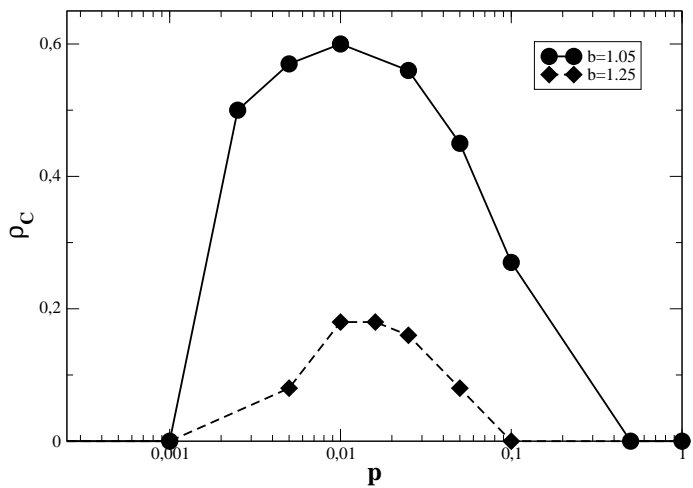}
\end{center}
\caption{Cooperator density $\rho_C$ [for temptation values of
  $b=1.05$ (circles) and $b=1.25$ (diamonds)] in the asymptotic state
  for LASW networks with $m=1$ and $N=400$ as a function of the
  link-adding probability $p$. All agents use the REP update
  rule. Note the logarithmic scale of the $x$ axis.}
\label{ultima}
\end{figure}

\begin{figure}
\begin{center}
\includegraphics[angle=0,width=10cm,clip]{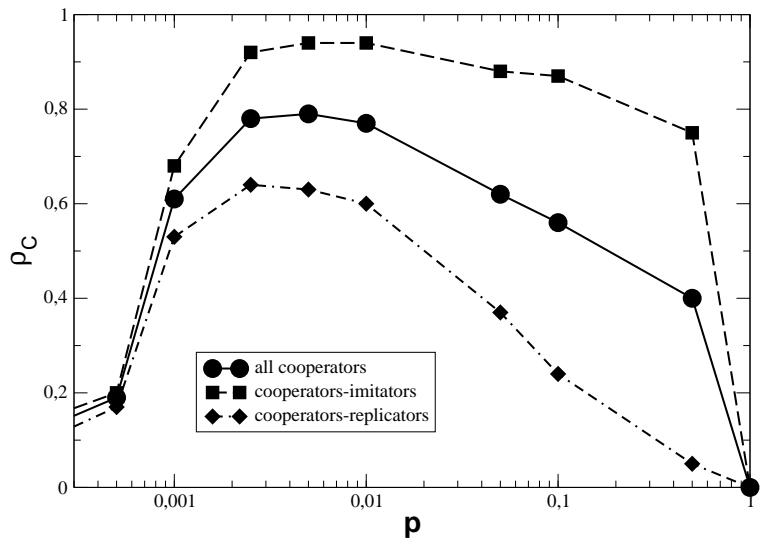}
\end{center}
\caption{Cooperator density $\rho_C$ (for a temptation value of
  $b=1.05$) in the asymptotic state for LASW networks with $m=1$ and
  $N=1000$ as a function of the link-adding probability $p$.  The
  system includes the two update strategies (half of the agents use
  the UI rule, half the REP one). The update rules did not evolve in
  time.}
\label{graf7}
\end{figure}

Let us begin from the case of a one-dimensional ring, that is, the
case $p=0$, starting from an initial configuration with fifty percent
of cooperators, $m=1$ and with UI as the update rule.  Numerical
simulations show that the final cooperator density is around $0.14$
for each value of $b\in(1,2]$ (we will give an analytical explanation
  for this value in Sec.\ \ref{intrprt}). On the other hand, in the
  opposite limit, $p=1$, the topology corresponds to a fully connected
  network, and thus there are no cooperators in the final
  configuration (see Section~\ref{mdl}).  Far from these two limits,
  highly non-trivial behaviour arises for intermediate values of the
  link-adding probability $p$. In Fig.~\ref{lak1} we plot the final
  density of cooperators on a system with UI dynamics, size $N=1000$,
  temptation $b=1.05$ and $1.25$, and initial cooperator density equal
  to $0.5$.  The results are really intriguing: For $p\in[0.002,0.2]$
  ($b=1.05$), and for $p\in[0.002,0.02]$ ($b=1.25$) there is a large
  plateau where cooperation is dramatically enhanced with respect to
  both the one-dimensional ring ($p=0$) and, of course, the complete
  graph ($p=1$). Such plateaus suggest the existence of an optimal
  region of the link-adding probability starting at $p\simeq p^*$. As
  observed from Fig.~\ref{lak1} the length of the optimal region
  decreases with $b$, as is the case with the cooperator density as
  well.  It is noticeable that the same picture emerges when the
  update rule used in the system is the REP one, as clearly shown in
  Fig.~\ref{ultima}: In this case we also observe an optimal region of
  the link-adding probability (starting at $p\simeq p^*$) in which
  cooperation is very large. This means that, apart from the
  (expected, \cite{roc09,per9}) smaller levels of cooperation in each
  stage with respect to the UI system, the fundamental mechanism
  underlying such phenomenon cannot be due to the update rule chosen,
  but to the topological features of the network. To further support
  this claim, we have verified that when update rules coexist in the
  population, the phenomenology remains practically unchanged, as
  shown in Figure~\ref{graf7}. Interestingly, even in such mixed
  dynamics UI rule enhances cooperation more than REP one, coherently
  with the results presented in the literature mentioned
  above. Finally, we note also that the cooperation levels of the two
  fractions of the population are similar to those observed when the
  corresponding rule is the only one used, providing further evidence
  of the independence of the results of the update rule considered. It
  is thus clear that the existence of an optimal range of short-cut
  density is due to structural features and thus it should be
  explained in topological terms.

\begin{figure}
\begin{center}
\includegraphics[angle=0,width=10cm,clip]{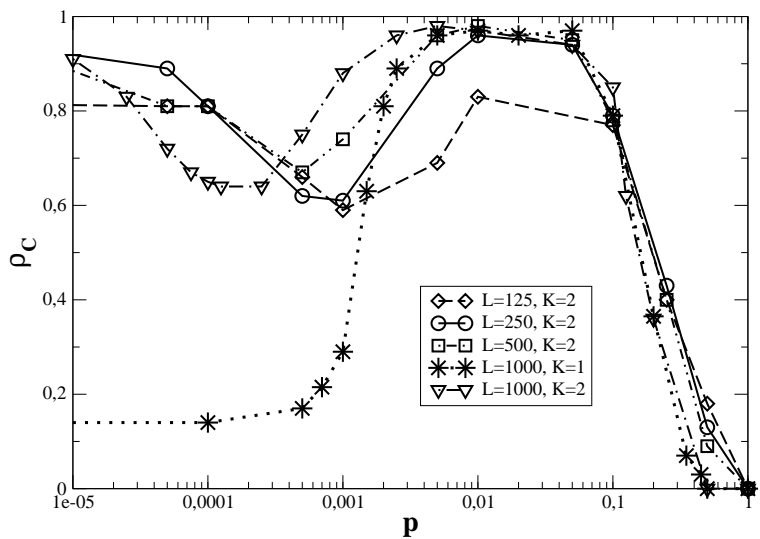}
\end{center}
\caption{Cooperator density $\rho_C$ (for a temptation value of
  $b=1.05$) in the asymptotic state of LASW networks with $m=2$ and
  $N=1000$ as a function of the link-adding probability $p$.  All
  agents use the UI update rule. Note the logarithmic scale of the $x$
  axis.}
\label{graf8}
\end{figure}

The above results correspond to the case $m=1$, which as explained
above we expect to be representative of the evolutionary outcomes for
larger values of the initial degree at least for $p>p^*$. This is
indeed verified in Fig.\ \ref{graf8}, where results for $m=2$ and
different systems sizes are shown. Our numerics confirm the existence
of a similar transition at $p\simeq p^*=2m/N$, where a plateau of
large cooperation densities is observed until this magnitude begins to
decrease as $p\rightarrow1^-$. It should be noted, however, that there
is an interesting (but not unexpected) difference with respect to the
case $m=1$ that arises from the behaviour for $p\rightarrow0^-$. In
this limit, the cooperative behaviour in the frozen state is very high
but, as we will see in the next Section, this is due to the different
geometrical properties of an euclidean ring with
$m\geq2$. Notwithstanding this special feature, the main conclusion of
Fig.\ \ref{graf8} is that for $p\gtrsim p^*$ the behaviour of the
system is independent of $m$, in the same way of the topology itself
of the system.

\begin{figure}
\begin{center}
\includegraphics[angle=0,width=10cm,clip]{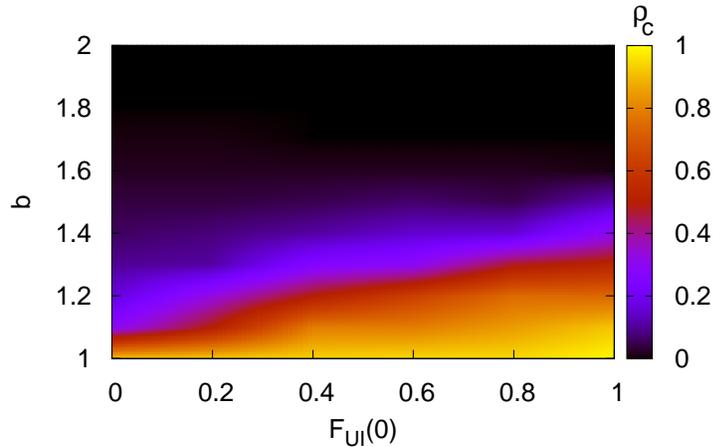}
\end{center}
\caption{The cooperator density $\rho_C$ in the asymptotic state for
  LASW networks with $m=1$, $N=2000$ and $p=0.01$, as a function of
  the temptation $b$ and of the initial density $F_{ui}(0)$ of
  imitators in the initial stage, is shown by means of a color scale.
  Update rules (UI and REP) coevolve with the strategies as described
  in the text; the initial density of cooperators is equal to
  $0.5$. Note that the larger $F_{ui}(0)$, the larger the asymptotic
  level of cooperation.}
\label{pag1}
\end{figure}

Finally, in order to present a more complete picture, we have also
analyzed the case in which strategies and strategy update rules
co-evolve, a line of work that has attracted much interest recently
\cite{per09}. We will consider the approach proposed in \cite{moya},
in which agents that decide to copy the strategy of a neighbour copy
her strategy update rule as well. Recent results \cite{cardillo} show
that such a co-evolutionary process gives rise to different results
depending on the topologies and the rules that are initially present
in the population. We have therefore carried out simulations with both
strategy update rules present in the population in different initial
fractions, and allowing them to be adopted as described
above. Figure~\ref{pag1} shows that even with evolving update rules,
around $p\simeq0.01$ cooperation is once again promoted, for values up
to about $b\simeq1.4$. This is a further hint on the robustness of the
cooperation enhancement process due to the short-cuts. In accordance
with the results above, we note that the range in which the
cooperation is enhanced decreases with the density of initial
replicators.  Therefore, all our results show that there are some
values of $p$ (and thus of the clustering coefficient, $-$ see
Section~\ref{swn}), in which cooperative behaviour is largely enhanced
regardless of the initial degree of the network and of the
evolutionary dynamics. In the next Section we give some theoretical
explanations of this phenomenon, which certainly plays an important
role in the promotion of cooperation on complex networks.

\section{Discussion}
\label{intrprt}

\subsection{One-dimensional ring}

Let us start our analysis from the simplest case: the one-dimensional
ring with $m=1$ ({\em i.e.}  the case with two neighbours, one left
and one right, per site), in which agents play PDG with UI as
microscopical update rule. We will begin by showing that the
asymptotic cooperation density does not depend on $b$. Having
$b\in(1,2]$, this is an immediate consequence of the remarkable
  property of the dynamics that a defector never changes strategy:
  Indeed, a defector surrounded by at least one cooperator earns at
  each game a payoff $\Pi\geq b>1$, whilst a cooperator surrounded by
  at least one defector earns a payoff $\tilde{\Pi}\leq1<\Pi$, so that
  a $D$-strategist will never find a better performing neighbour with
  $C$ strategy. Thus, the evolution of the system is independent of
  the value of $b$, provided $b>1$. Furthermore, it follows that the
  cooperator density (that is, the average number of cooperators over
  the total number of agents) is a decreasing function of time:
  $\rho^{\infty}_C\leq\rho_0$, where
  $\rho^{\infty}_C=\lim_{t\rightarrow+\infty}\rho_C(t)$ and
  $\rho_0=\rho_C(t=0)$ are the final and initial cooperator densities
  respectively.

As a second result, we will now show that a necessary condition for a
cooperator not to change strategy is to be set in the middle of a
cluster of at least three consecutive cooperators. To this end, let us
note that a configuration of the form $DDCCCDD$ with cooperators in
the boundaries is stable, because the payoffs of the $7$ players are
respectively $b, b, 1, 2, 1, b, b$. Thus, the defectors and the
central cooperator do not have any reason to change, while the
cooperators in the boundary of their cluster imitate the central one,
and hence they keep the $C$ strategy. Nevertheless, reasoning as
above, it must be noticed that a configuration $DCCCDC$ (with
cooperators in the boundary) evolves to $DCCDDD$ and finally to
$DDDDDD$. In general, isolated defectors give birth to a triplet of
$D$-strategists: $\dots CCCCDCCCC\dots$ goes to $\dots
CCCDDDCCC\dots$, which is a stable configuration.

Having the above considerations in mind, we can say that the final
cooperator density must be proportional to the density of clusters of
at least three consecutive cooperators in the initial state, minus the
effect of the initially isolated defectors $\alpha(\rho_0)$. Then, we
can write down
\begin{equation}
\label{rhofin1}
\rho_C=3(1-\rho_0)\rho_0^3(1-\rho_0)+4(1-\rho_0)\rho_0^4(1-\rho_0)+\dots-\alpha(\rho_0)\ ,
\end{equation}
where each term $l(1-\rho_0)\rho_0^l(1-\rho_0)$ is the contribution of initial clusters of
$l$ cooperators to $\rho_C$ (that is, the number of cooperators $l$ in the cluster times the probability
to find one defector, then $l$ consecutive cooperators, finally another defector in the initial configuration).
Besides, we have $\alpha(x)\in[0,1)$ and $\lim_{x\rightarrow0^+}\alpha(x)=\lim_{x\rightarrow1^-}\alpha(x)=0$.
The previous equation can be rewritten as
\begin{equation}
\label{rhofin2}
\rho_C=\rho_0^3(1-\rho_0)^2\sum_{j=0}^{+\infty}(j+3)\rho_0^j-\alpha(\rho_0)=\rho_0^3\cdot(3-2\rho_0)-\alpha(\rho_0)\ .
\end{equation}

We stress the fact that the infinite sum in equations~(\ref{rhofin1})
and~(\ref{rhofin2}) does converge correctly to 1 for $\rho_0\rightarrow1^-$,
as can be easily seen in the last member of the~(\ref{rhofin2}).

As in our simulations we always started from $\rho_0=0.5$, we have to
evaluate the correction term $\alpha(\rho_0=0.5)$. This can be
cumbersome, but the most important correction around $\rho_0\simeq0.5$
is that arising from configurations of the form $\dots DCC\dots$
$CCDCC\dots CCD\dots$, {\em i.e.} two clusters of consecutive
$C$-strategists being made up by at least three cooperators. Each of
such initial configurations leads to the removal of the two
cooperators surrounding the central defector. In fact the effect is
somewhat larger, because some clusters with 3 cooperators are removed
when they are close to a larger cooperator cluster as explained above.
Ignoring this effect as a first approximation we can compute, analogously
to the previous calculation
\begin{equation}
\label{alfa}
\alpha(\rho_0)\simeq3P_3(1-\rho_0)P_3=\frac{3\rho_0^6}{1-\rho_0}\ ,
\end{equation}
with
\[
P_3=(1-\rho_0)^2\cdot(\rho_0^3+\rho_0^4+\rho_0^5+\dots)=\rho_0^3(1-\rho_0)^2\sum_{j=0}^{+\infty}\rho_0^j=
\frac{\rho_0^3}{1-\rho_0}\ .
\]
Then, in case of one-dimensional rings with degree equal to 2
and $\rho_0=0.5$, from equations (\ref{rhofin2}) and (\ref{alfa}) we
find $\rho_C\simeq0.156$, with an $8\%$ error respect to the real
value of about $0.143$. This good (even though imperfect) agreement shows that we have indeed
identified the main mechanisms governing the evolution in the
one-dimensional ring.

Note also that the general behaviour described above
(cooperator density decreasing function of time, with final value only
depending on the initial configuration) holds for rules
such as UI and REP, where there is no possibility of making mistakes,
{\em i.e.} one node never copies the strategy of a neighbour with
smaller payoff. In this context, it is worth noticing that the model
discussed in this subsection is very close to the one described by
Eshel and coworkers \cite{esh98}, where mistakes are possible and
defectors (called "egoists" in the paper by Eshel {\it et al.}) can at
some point become cooperators ("altruists").

\subsection{The role of short-cuts}

Let us now consider the case of $p>0$ (always with $m=1$). It is easy
to see that the presence of short-cuts changes in general the
stability of a configuration. In particular, it is no longer true that
$\rho^{\infty}_C\leq\rho_0$.  Indeed, think of a long cluster of
defectors, a configuration which is stable in a ring. If we connect
one of the internal agents with a far away cooperator surrounded by
two other $C$-strategists, at the subsequent step the long-range
connected defector will flip toward cooperation.  More generally, the
fact that a defector can be directly connected with more than two
cooperators makes the transition $D\rightarrow C$ possible, so that
$\rho^{\infty}_C$ is expected to increase with $p$ increasing, at
least for small values of the link adding probability. On the other
hand, in the limit $p\rightarrow1^-$ we know that the cooperation
level must vanish~\cite{moya}. Therefore, we expect a non trivial
behaviour with a maximum of cooperation at some value
$\tilde{p}\in(0,1)$. Actually, as shown in Figure~\ref{lak1}, in
numerical simulations we see a relatively long plateau with a great
level of cooperation around $p\sim0.01$.

Such behaviour can be understood by means of topological
considerations. As pointed out in Section~\ref{swn}, our network
undergoes a sort of (smooth) transition at $p\sim p^*=2k/L$. Indeed,
for smaller values of $p$ the system is topologically equivalent to a
WS small-world network, whilst for larger values it behaves like an ER
one. When we consider the system in Figure~\ref{lak1}, we have
$p^*\simeq0.002$, and we correspondingly see a sudden increase of the
final cooperation just around this value itself. Moreover, in the WS
regime the number of short-cuts is not enough yet to affect
appreciably the dynamics of the system (there is only a very slight
increase of the final cooperators due to the effect described just
above). However, when this number becomes large enough (more
precisely, when the number of the sort-cuts is larger than that of
regular links), the presence of several hubs connected with stable
cooperators makes the configuration of the system very favorable to
the final emergence of the cooperation. Finally, when $p$ approaches
$1$, the number of short-cuts becomes so large that the system is
practically a fully connected graph, hubs are no longer hubs but they
are almost like any other node, and the cooperation is completely
suppressed~\cite{moya}.

\begin{figure}[h]
\begin{center}
\includegraphics[angle=0,width=8.5cm,clip]{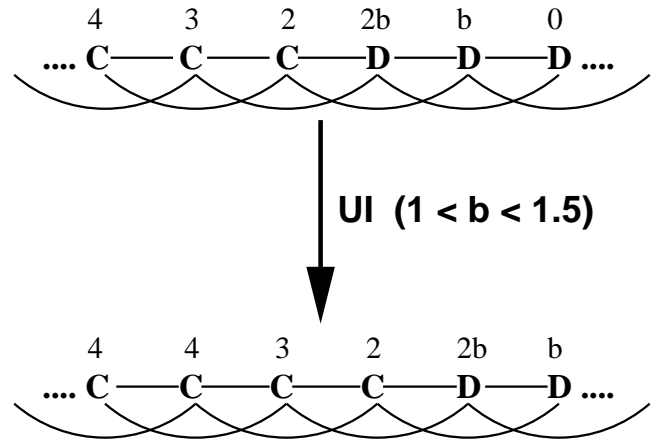}
\end{center}
\caption{Sample of elementary dynamics on the boundary between two
  long clusters of different strategy ($m=2$); the numbers above the
  individuals (singled out by their strategy) are the payoffs earned
  after just that time step.  With the UI microscopical update rule
  and $b\in(1,3/2)$, the cooperating cluster gains one position at
  each elementary time step.}
\label{ccdd}
\end{figure}

Similar effects happen for $m\geq2$, but in that case everything is
much more complicated. In any event, in the one dimensional case
($p=0$) the dynamics is again purely deterministic (with the UI update
rule), and the final frozen state depends only on the initial
configuration of the system. Given the random initial conditions we
are using, the cooperator density of the frozen state is around
$0.96$, for large enough system sizes. This means that, for an
euclidean ring, increasing the coordination number enhances enormously
the cooperation.  This could be expected because, following the
reasoning illustrated above, with 4 neighbours, a cooperator is likely
to be imitated by a defector. For example, let us consider the border
between two long clusters, each of opposite strategy (see
Figure~\ref{ccdd}). Here, provided that $b\in(1,3/2)$, the first
defector is connected with a cooperator with larger payoff, so that
she will flip at the next stage to the $C$ strategy (whilst no
cooperator is connected with an analogous better-fitting
defector). Consequently, the cooperator cluster will increase its size
in one unit at each time step, until a frozen state almost completely
full of cooperators is reached. In case the dynamics was given
by the REP rule, the result is the same, although the time scale
needed to reach the asymptotic scale is much longer due to the
probabilistic nature of the rule.

When short-cuts are added, we have again a different regime, depending
on whether $p<p^*$ or $p>p^*$. In the first regime, there are less
short-cuts than regular links, and hence the system is topologically
like a WS small-world network with $m=2$, {\em i.e.} globally it
behaves like a random network (very low diameter) whilst locally it
resembles a regular lattice (clustering coefficient almost constant
around $\chi_0(m=2)=1/2$, which decreases reaching a minimum only for
$p\rightarrow p^*$).  When a few short-cuts are present in the system,
defection is initially enhanced for the same mechanism that promotes
cooperation for $m=1$: A short-cut could connect a cooperator with a
defector with larger payoff, driving the first to change her strategy
to defection. Finally, once that $p>p^*$, the system becomes
equivalent to an ER random network and the cooperation is enhanced
again, until for $p$ large enough we end up with a complete graph
where only defectors survive. As can be seen in Figure~\ref{graf8},
for each value of $N$, and hence of $p^*=2m/N$, the final level of
cooperation decreases when $p\lesssim p^*$, then increases rapidly as
soon as $p\gtrsim p^*$, before vanishing completely for
$p\rightarrow1^-$, confirming the interpretation given above.

It must be noticed that all the reasoning and considerations we have
done in this section assume that the microscopical update rule is the
unconditional imitation. We have also indicated along the way that
choosing the REP rule will only change the time scale of the
evolution. In addition, in so far as there is no possibility of
mistakes, the general scheme proposed above also holds with mixed
update rules (see Figures~\ref{graf7} and~\ref{pag1}): Cooperation is
mostly suppressed for $p\lesssim p^*$ (WS topology), greatly enhanced
for $p\gtrsim p^*$ (ER topology), and totally suppressed for
$p\rightarrow1^-$ (fully connected graph topology). The robustness of
such scheme with respect to the update rules adopted by the individuals
confirms the fundamental role of the short-cuts for the ultimate
fate of the dynamics.

\section{Conclusions and perspectives}
\label{cncl}

In this paper we have investigated how the topology of a system in
which a population of many individuals interact by means of the PDG
can influence the emergence of stable cooperative behaviours, focusing
on the effect of short-cuts. For this purpose we exploited a simple
model which can assume different topological features of different
systems (euclidean lattice, WS small-world network, ER random network,
and fully connected graph) by tuning only the value of a single
parameter $p\in[0,1]$ (the link-adding probability). In this way, it
is easier to distinguish the precise role each particular topology
plays in enhancing (or suppressing) cooperation. In particular, we
have shown how a regular euclidean lattice promotes cooperation only
for $m\geq2$, $m$ being the coordination number of a one-dimensional
ring, while for $m=1$ only a small minority of cooperators survives in
the final frozen state. We have also seen that the WS topology does
not enhance cooperation, whilst the ER network configuration is the
best cooperation promoter of all in this family. Finally, as was
already well known, in a fully connected graph (or, equivalently, in
the mean-field approximation) only defectors can survive in the final
state.

Since the only free parameter of the network model is the link-adding
probability $p$ and the results do not depend on the particular
dynamics considered, it is the density of short-cuts actually existing
in the system that must determine the ultimate fate of the dynamics.
Indeed, when $0<p\lesssim p^*$, the presence of short-cuts (whose
number is still smaller than that of regular links) causes at most a
perturbation of the configuration emerging in the euclidean case (the
final level of cooperation is very slightly increased if $m=1$, not
dramatically decreased otherwise), while for $p\gtrsim p^*$ in the
network many hubs appear, and a cooperator hub is very likely to act
as a seed for the spreading of the cooperation throughout the whole
system. By means of such a mechanism, the cooperation is enhanced much
more than in the other topological regimes. Finally, when
$p\rightarrow1^-$, every site ends up being connected with all the
others and the system becomes a fully connected graph, where only
defectors can survive.  

The picture that emerges from our research (that reached the limit 
of the fully connected network) and from previous works \cite{masu,san05,fu07,du09,vuk08}
is that the ER
topology may be the best one among homogeneous or mildly heterogeneous
graphs to promote
cooperation, and that this property is quite robust, given that the
same (qualitatively) behaviour can be found with different update
rules (UI and REP dynamics), and also mixing them and letting them
co-evolve in the same system. We want to stress that this mechanism
for the promotion of cooperation has nothing to do with the one based
on clustering discussed in \cite{roc09,per9}, as the clustering of the
networks where we observe a plateau in the cooperation level is very
small ($\sim p$). This is hence a novel mechanism that has not been
discussed before and that confers to short-cuts a very important role
in the emergence of cooperation. In fact, they can be used to
eliminate the effect of topological traps discussed in \cite{traps},
which arises due to the existence of bottlenecks and lack of redundant
paths. The work summarized here confirms that a not so large fraction of
short-cuts can have dramatic effects on the final level of cooperation
achieved in the population and clarifies the origin of this mechanism. 
On the other hand, a more-in-depth study of the role of
other update rules is still needed; in particular, in the present work
we have neglected the family of rules allowing the agents to make some
mistakes, {\em i.e.}, allowing an agent to imitate a worse-fitting
neighbour, as for instance happens with the Moran
rule~\cite{per9,mor} or with the dynamics in \cite{vuk08}. 
The case of a two-dimensional initial lattice is
also interesting, although we envisage that in that case it will be
difficult, if not impossible, to gain insights from an analytical
viewpoint as we have done here. Future works should deal with these
issues.

\section*{Acknowledgments}
D.\ V.\ is supported in part by a postdoctoral contract from
Universidad Carlos III de Madrid.  A.\ S.\ was supported in part by
grants MOSAICO and Complexity-NET RESINEE (Ministerio de Ciencia e
Innovaci\'on, Spain) and MODELICO-CM (Comunidad de Madrid,
Spain). J.\ G.-G. is supported by the MICINN through the Ram\'on y
Cajal program and grants FIS2008-01240 and MTM2009-13838.

\end{document}